\documentclass[aps,prd,preprint]{revtex4-1}
\usepackage {graphicx}
\usepackage[toc,page]{appendix}
\usepackage{amsmath}
\usepackage{tikz}
\newcommand{\be}{\begin{equation}}
\newcommand{\ee}{\end{equation}}
\newcommand{\bea}{\begin{eqnarray}}
\newcommand{\eea}{\end{eqnarray}}
\begin{document}
\title{The radiation zone in general relativity}
\author{David Garfinkle}
\email{garfinkl@oakland.edu}
\affiliation{Dept. of Physics, Oakland University, Rochester, MI 48309, USA}
\affiliation{Leinweber Center for Theoretical Physics, Randall Laboratory of Physics, University of Michigan, Ann Arbor, MI 48109-1120, USA}

\date{\today}

\begin{abstract}
The radiation zone in electrodynamics is the region far enough away from the charges that the $1/r$ part of the field dominates over the $1/{r^2}$ piece.  This concept is key in explaining two puzzling aspects of general relativity: The first is an old paradox that invokes the equivalence principle to argue that a static charge in a gravitational field will radiate. The second is the fact that while there are astrophysical sources of gravitational radiation, we do not have any man-made sources.

\end{abstract}


\maketitle


\section{Introduction}

The electromagnetic field of a set of moving charges can be thought of as consisting of a Coulomb piece that falls off as $1/{r^2}$ and a radiation piece that falls off as $1/r$.  In order to make unequivocal statements about radiation, one must be in the radiation zone: that is far enough out that the radiation piece is dominant.\cite{Griffiths}  For an assembly of charges that oscillates with a definite angular frequency $\omega$, the radiation zone can be characterized in term of the associated wavelength, which is $\lambda = 2 \pi c /\omega$ where $c$ is the speed of light.  Here the radiation zone is simply characterized by $r \gg \lambda$.  For a single charge with acceleration $a$ an examination of the $1/r$ and $1/{r^2}$ pieces of the electromagnetic field yields the conclusion that the radiation zone is characterized by $r \gg {c^2}/a$.

These are all well known facts about electrodynamics, but they can also be used to illuminate two puzzling features of general relativity: one old and one new.

The old one is a paradox in which one invokes the principle of equivalence to argue that a charge at rest in a gravitational field will radiate.  

The new one is the fact that though we have detected astrophysical sources of gravitational waves, we have no artificial sources strong enough to be detected.

We will treat each of these puzzles in turn.

\section{The charge on a table paradox}

What is the electromagnetic field of a charge on a table on the surface of the Earth?  We all know the answer.  To the extent that we neglect the effects of general relativity, it is just the Coulomb field of the charge.  Even taking general relativity into account, the electromagnetic field of the charge is a static solution of Maxwell's equations in the curved spacetime of the Earth's gravitational field that is almost exactly equal to the Coulomb field.  

However, we also know that the equivalence principle states that static position in a uniform gravitational field is equivalent to constant acceleration in the absence of a gravitational field.  Since accelerating charges radiate, it seems that the equivalence principle gives us reason to believe that the charge on the table is radiating!

This paradox has a surprisingly long history.{\cite{Pauli,Schott,Rohr,Feynman}}  We all know that the charge doesn't radiate.  But then what is wrong with the argument based on the equivalence principle?  Since the argument makes use of both the concept of radiation and of the equivalence principle, we will start by considering when each of these concepts can be applied.

  For a single point charge undergoing acceleration $a$ the radiation zone is given by $r \gg {c^2}/a$.  It is instructive to see what this gives for the charge on the tabletop on the surface of the Earth.  Since the equivalence principle instructs us to think of this as a charge undergoing acceleration $g=9.8 {\rm m}/{{\rm s}^2}$ and since $c=3.0 \times {{10}^8} {\rm m}/{\rm s}$, we find that to be in the radiation zone we must be at a distance at least of order ${10}^{16} {\rm m}$ away from the source.  

The equivalence principle considers only a uniform gravitational field.  Thus, in order to use it, all the phenomena that we are considering must take place in a region that is small enough that the gravitational field can be approximated as uniform.  For the gravitational field of the Earth, this means that in order to use the equivalence principle we must consider a region small compared to the size of the Earth, or in other words a region whose radius is small compared to the radius of the Earth, which is approximately $6.37 \times {{10}^6} {\rm m}$.  

Thus in order to even set up the paradox of the charge on a tabletop on Earth we would need to consider the properties of the charge's electromagnetic field at a distance that is large compared to ${{10}^{16}} {\rm m}$ but small compared to ${{10}^7} {\rm m}$.  Needless to say, there is no such distance.  But that also means that the paradox cannot be set up: for the Earth, the charge on the tabletop paradox does not occur.

Nonetheless, since the paradox is a question of principle, then even though it doesn't occur on Earth, we might wonder whether there is some gravitating body on which it might occur.  That is: consider a body with mass $M$ and radius $R$.  Are there values of $M$ and $R$ for which the paradox occurs? For such a body, the acceleration at its surface would be $GM/{R^2}$ and so the size of the radiation zone would be ${c^2}{R^2}/(GM)$.  In order for something to be both in the radiation zone and at a distance small compared to the size of the body, we would need the size of the radiation zone to be small compared to the size of the body.  That is we would need
\be
{\frac {{R^2}{c^2}} {GM}} \ll R
\ee
which is equivalent to 
\be
R \ll {\frac {GM} {c^2}}
\ee
Or in other words the object would need to be much smaller than its Schwarzschild radius.  But this can't occur because any such object would not be static: it would have already collapsed to form a black hole.

\section{why are there no artificial sources of gravitational radiation?}

It is now a little more than a decade since the first direct detection of gravitational radiation.\cite{LIGO}  Because gravity is such a weak force, the detectors need to be huge laser interferometers whose incredible precision and insulation from noise requires heroic feats of engineering and experimental physics. 

Gravitational waves are perturbations $h_{ij}$ of the spacetime metric, and lead to fractional changes in the size of the interferometer arms of order $h_{ij}$.  The first detection of gravitational waves, done by LIGO\cite{LIGO}, had ${h_{ij}} \simeq {{10}^{-21}}$.

In addition, because gravity is such a weak force, only the strongest sources of gravitational radiation can be detected.  Just as electromagnetic waves are made by accelerating charges, so gravitational waves are made by accelerating masses.  The strongest sources are binary systems of two black holes in orbit around their common center of mass.  As the black holes orbit, they emit gravitational radiation, lose energy thus shrinking the size of the orbit, until finally they merge to form a single larger black hole and a final burst of gravitational radiation.

Let's look more closely at the claim that only the strongest sources can be detected.  The binary black hole systems detected by LIGO are intrinsically strong, but they are also very far away.  In particular the first detection was of a system at a distance of about a billion light years.  Since a radiation field falls off as $1/r$ a weaker source at a smaller distance can give the same signal. Since there are several precision laboratory scale measurements and tests of gravity,\cite{adelberger} why couldn't we do a laboratory scale production and detection of gravitational waves?  (i.e. a sort of gravitational analog of the Hertz experiment for electromagnetic waves).

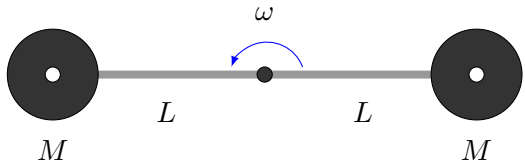
\begin{figure}
\centering
\begin{tikzpicture}
\draw[line width=3pt, gray!80] (-3,0) -- (3,0);
\foreach \x in {-2.8, 2.8} {
\draw[fill=black!80] (\x,0) circle (0.6);
\draw[fill=white] (\x,0) circle (0.1);
}
\draw[fill=black!80] (0,0) circle (0.1);
\node at (-2.8,-1) {$M$};
\node at (2.8,-1) {$M$};
\node at (1.3,-0.5) {$L$};
\node at (-1.3,-0.5) {$L$};
\draw[-latex,blue] (0.5,0.1) arc
[
start angle=20,
end angle=160,
x radius = 0.5,
y radius =0.5
];
\node at (0.0,0.8) {$\omega$};
\end{tikzpicture}
\caption{rotating barbell consisting of two masses $M$, with separation $2L$, rotating around the center of the bar at angular frequency $\omega$.}
\label{fig1}
\end{figure}

To focus on a concrete example, let's make our source a rotating barbell, as shown in figure (\ref{fig1}).
Since we want the source to be as strong as we can make it, we want the rotational angular velocity to be as high as we can make it.  Fast ultracentrifuges rotate at a rate of about $1.5 \times {{10}^5} {\rm rpm}$ which corresponds to an angular velocity of about 
$(\pi /2) \times {{10}^4} {{\rm s}^{-1}}$.  We will see later that the angular frequency of the gravitational waves generated by the rotating barbell is twice the angular frequency of rotation.  This gives rise to a wavelength of the gravitational waves of
\be
\lambda = 6.0 \times {{10}^4} {\rm m}
\ee
Thus the wave zone is given by $r \gg 60 {\rm km}$.

This provides a partial answer to the question of a laboratory size gravitational Hertz experiment: we can't have a laboratory size gravitational Hertz experiment because the wave zone is much larger than the size of the lab!  Nonetheless, this actually simplifies the experimental design:  Let's build our laboratory size source a few hundred kilometers from one of the LIGO detectors and invite our friends at the LIGO laboratory to detect it.  How does its strength compare to that of the black hole signals detected by LIGO?  We now turn to that part of the calculation.  

In electrodynamics, for sources moving much slower than light the vector potential in the radiation zone is given by the expression\cite{Griffiths}
\be
{\vec A}({\vec r},t) = {\frac {\mu _0} {4 \pi}} {{{\dot {\vec p}}({t_r})} r}
\label{larmour}
\ee
Here $\vec p$ is the electric dipole moment of the source, and the retarded time $t_r$ is given by ${t_r}=t-r/c$.

Though the equations of general relativity can be fearsomely complicated, in the weak field slow motion limit they are actually very similar to Maxwell's equations. The analog for general relativity of eqn. (\ref{larmour}) is that in the radiation zone for sources moving slowly compared to the speed of light, the metric perturbation $h_{ij}$ is given by\cite{flanaganhughes}  
\be
{h_{ij}}({\vec r},t) = {\frac {2G} {c^4}} {\frac {{{\ddot Q}_{ij} ^{TT}}({t_r})} r}
\label{hamplitude}
\ee
Here ${Q_{ij}}$ is the mass quadrupole moment of the source, which is the trace-free part of the moment of inertia tensor.  The superscript $TT$ stands for ``transverse-traceless.'' Here transverse denotes that the quantity is projected to be orthogonal to the radial direction (because gravitational waves are transverse, like electromagnetic waves). Also trace-free means that in addition the quantity is projected to have zero trace (because gravitational tidal forces distort shapes but do not change the volume).  

For a set of masses $m_\alpha$ at positions ${\vec x}_{\alpha}$ The moment of inertia tensor $I_{ij}$ is given by
\be
{I_{ij}} = {\sum _\alpha} {m_\alpha} {x^i _\alpha}{x^j _\alpha}
\label{inertia}
\ee
Now consider a barbell of length $2L$, with a mass $M$ at either end, and rotating around its center at angular velocity $\omega$.  We will pick the origin to be the center of the barbell, and the $xy$ plane to be the plane of rotation.  Then from eqn. (\ref{inertia}) it follows that
\be
{I_{ij}}({t_r}) = M {L^2} \left ( 
\begin{matrix} {\cos^2} \omega {t_r} & \cos \omega {t_r} \sin \omega {t_r} & 0 \\
 \cos \omega {t_r} \sin \omega {t_r} &  {\sin^2} \omega {t_r} & 0 \\ 0 & 0 & 0
\end{matrix}
\right )
\ee
We then find
\be
{{\ddot I}_{ij}}({t_r}) = 2 M {L^2}{\omega ^2} \left ( 
\begin{matrix} \cos 2 \omega {t_r} & \sin 2\omega {t_r} & 0 \\
 \sin 2 \omega {t_r} &  - \cos 2 \omega {t_r} & 0 \\ 0 & 0 & 0
\end{matrix}
\right )
\label{ddotQ}
\ee
Since ${\ddot I}_{ij}$ is trace-free, it is also equal to ${\ddot Q}_{ij}$.
Since $\omega = 2 \pi c /\lambda$, we have
\be
2 M {L^2} {\omega ^2} = 8 {\pi ^2} M {c^2} {{\left ( {\frac L \lambda} \right ) }^2}
\label{amplitude}
\ee
Finally, using eqns. (\ref{ddotQ}) and (\ref{amplitude}) in eqn. (\ref{hamplitude}) we find the following estimate for the order of the components of $h_{ij}$.
\be
{h_{ij}} \simeq 8 {\pi ^2} {\frac {G M} {{c^2} r}} {{\left ( {\frac L \lambda} \right ) }^2}
\label{hamplitude2}
\ee
It is the factor of $G/{c^2}$ that is mainly responsible for the tiny size of the gravitational wave amplitude, since $G/{c^2} = 7.4 \times {{10}^{-28}} {\rm m}/{\rm kg}$.  However, there are additional small factors having to do with the fact that the radiation zone is much larger than the laboratory scale.  Let's take $r=5 \lambda = 300 {\rm km}$ to be safely in the wave zone.  Then we find
\be
{h_{ij}} \simeq 5 \times {{10}^{-41}} {\frac M {1 {\rm kg}}} {{\left ( {\frac L {1 {\rm m}}} \right ) }^2}
\label{hamplitude3}
\ee
This is about 20 orders of magnitude smaller than the astrophysical sources detected by LIGO!
\section*{Acknowledgments}

It is a pleasure to thank Chris Vuille, Tom Roman, and Yang Xia for helpful discussions.


\begin{thebibliography}{}

\bibitem{Griffiths}
D. Griffiths, {\it Introduction to Electrodynamics}, fourth edition, Cambridge University Press (2017)

		
\bibitem{Pauli} W. Pauli, ``Theory of Relativity'', P. 93, Pergamon, New York, 1958
		
\bibitem{Schott} G. A. Schott, Phil. Mag. 29, 49, 1915
		
\bibitem{Rohr} T. Fulton and F. Rohrlich, Ann. Phys. 9, 499, 1960

\bibitem{Feynman} R.P. Feynman, Feynman Lectures on Gravitation, Addison-Wesley, 1995

\bibitem{LIGO}
Abbott et al. Phys. Rev. Lett. {\bf 116} 061102 (2016)

\bibitem{adelberger}
J. Lee, E. Adelberger, T. Cook, S. Fleischer, and B. Heckel, Phys. Rev. Lett. {\bf 124} 101101 (2020)

\bibitem{flanaganhughes}
E. Flanagan and S. Hughes, New J. Phys. {\bf 7}, 204 (2005)


\end{thebibliography}
\end{document}